\documentclass{iau}
\usepackage{graphicx}

\newcommand{\Ha}{H$\alpha$}
\newcommand{\kms}{km~s$^{-1}$}
\newcommand{\dhline}{\hline & \\[-1.7em]\hline}
\newcommand{\NiI}{\hbox{{\rm Ni}\kern 0.1em{\sc i}}}
\newcommand{\FeI}{\hbox{{\rm Fe}\kern 0.1em{\sc i}}}

\title[Sequential Chromospheric Brightenings] 
{Relationships Between Sequential Chromospheric Brightening and the Corona}

\author[M. S. Kirk {\it et al.}]   
{Michael S. Kirk$^1$
 K. S. Balasubramaniam$^2$
 Jason Jackiewicz$^3$
 \and Holly R. Gilbert$^4$}

\affiliation{$^1$ NASA Goddard Space Flight Center, Catholic University of America \\ 8800 Greenbelt Rd.,
Code 670, Greenbelt, MD 20771, USA \\ email: {\tt michael.s.kirk@nasa.gov} \\[\affilskip]
$^2$ AFRL Battlespace Environment Division (AFRL/RVB) \\[\affilskip]
$^3$ Department of Astronomy, New Mexico State University \\ Las Cruces, NM 88003, USA \\[\affilskip]
$^4$ NASA Goddard Space Flight Center \\ 8800 Greenbelt Rd.,
Code 670, Greenbelt, MD 20771, USA}

\pubyear{2017}
\volume{xxx}  
\pagerange{xxx}
\jname{Title of IAU Symposium}
\editors{A.C. Editor, B.D. Editor \& C.E. Editor, eds.}
\begin{document}

\maketitle

\begin{abstract}
The chromosphere is a complex region that acts as an intermediary between the magnetic flux emergence in the photosphere and the magnetic features seen in the corona. Large eruptions in the chromosphere of flares and filaments are often accompanied by ejections of coronal mass off the sun. Several studies have observed fast-moving progressive trains of compact bright points (called Sequential Chromospheric Brightenings or SCBs) streaming away from chromospheric flares that also produce a coronal mass ejection (CME). In this work, we review studies of SCBs and search for commonalties between them. We place these findings into a larger context with contemporary chromospheric and coronal observations. SCBs are fleeting indicators of the solar atmospheric environment as it existed before their associated eruption. Since they appear at the very outset of a flare eruption, SCBs are good early indication of a CME measured in the chromosphere.


\end{abstract}

\firstsection 
\section{Introduction}
During the eruption of two ribbon chromospheric flares there is often a significant number of compact brightenings in concert with the flare eruption  when observed in \Ha. These chromospheric bright points are temporally associated with the eruption but spatially separated from the evolving flare ribbon. One class of these flare-adjacent chromospheric brightenings was first classified by \cite[Balasubramaniam et al. (2005)]{2005ApJ...630.1160B}. Using a multi-wavelength data set to analyze an M2.7 flare on 2002 December 19, \cite[Balasubramaniam et al. (2005)]{2005ApJ...630.1160B} observed a large scale coronal dimming, sympathetic flares in both the north and south hemispheres, and a halo coronal mass ejection (CME). In \Ha\ images of the same event, the loop eruption manifested itself as flare precursor-brightenings and co-spatial propagating chromospheric brightenings. Termed sequential chromospheric brightenings (SCBs), they appeared as distinctly individual propagating points of brightening. 

More than a decade has passed since SCBs were first observed in \Ha\ (6562.8~\AA) and analyzed by \cite[Balasubramaniam et al. (2005)]{2005ApJ...630.1160B}. In that time, several additional studies have been completed with significant progress toward understanding the driving forces behind these compact features. \cite[Kirk et al. (2013)]{2013SoPh..283...97K} refined the technique for detecting SCBs and developed a systematic mechanism for identifying and measuring properties of SCBs by employing an automated detection and tracking algorithm. They concluded that SCBs originate during the impulsive rise phase of the flare and often precede the \Ha\ flare intensity peak. \cite[Kirk et al. (2012a,b)]{2012ASPC..463..267K,2012ApJ...750..145K} discovered that the nature of SCBs are phenomenologically distinct from other compact brighteings observed in the chromosphere due to their impulsive intensity signatures, unique Doppler velocity profiles, and origin in the impulsive phase of flare's \Ha\ intensity evolution. As an ensemble, SCBs are tracked to propagate outward, away from the flare center, at velocities on the order of $50$~\kms~(\cite[Kirk et al. 2012a]{2012ASPC..463..267K}).

\begin{table}
  \begin{center}
  \caption{A summary of the events used to investigate the evolution of SCBs. The time listed is the start time of the flare or filament eruption in \Ha. The Heliographic Stonyhurst (HGS) coordinates of approximate centroid of each event is listed for reference. Propagation velocities without stated uncertainties were marginal with a high amount of marginal SCB detections. The each was studied in detail by \cite[Balasubramaniam et al. (2005)]{2005ApJ...630.1160B}, \cite[Balasubramaniam et al. (2006)]{2006ilws.conf...65B}, \cite[Kirk et al. (2012a)]{2012ASPC..463..267K},  \cite[Kirk et al. (2012b)]{2012ApJ...750..145K}, \cite[Kirk et al. (2014) ]{2014ApJ...796...78K}, or \cite[Kirk et al. (2017)]{Kirk2017} and are labeled {\bf Ba}, {\bf Bb}, {\bf Ka}, {\bf Kb}, {\bf Kc} or {\bf Kd} respectively.}
  \label{tab1}
 {
  \begin{tabular}{llllllc}\dhline 
Event & Time & HGS & GOES & Visual & SCB Propagation & Previously  \\ 
Date  & UT & Coordinates & Class & CME &(\kms) & Studied  \\
\hline
2002-12-13 & 17:10    &	$18^{\circ}$ E,	$33^{\circ}$ N& C6.8 & yes & $65\pm 5$  & Ba \\
&&&&&&\\
2002-12-19 & 21:34 & $9^{\circ}$ W, $15^{\circ}$ N & M2.7 & yes &  $84.8\pm 9.7$ & Ba, Kb \& Kd  \\
		&		&	  					&	&	&$260\pm 19.4$ &	\\
2003-01-24 &  22:04 & $ 2^{\circ}$ E, $20^{\circ}$ S& C1.3 & yes & 100 & Ba \\
&&&&&&\\
2003-02-06 &  16:30  & $43^{\circ}$ W, $5^{\circ}$ N& None & yes & 80 & Ba \\ 
&&&&&&\\
2003-03-06 & 15:08 & $0^{\circ}$ W, $27^{\circ}$ N & None & no & $220\pm143$ & Kd  \\
&&&&&&\\
2003-05-09 & 15:18 & $1^{\circ}$ E, $35^{\circ}$ N & B6.6 & yes & $73.9\pm 10.9$ & Bb \& Kd  \\
		&		&	  					&	&	& -$8.5\pm 3.6$ &	\\
2003-06-11 & 17:27 & $23^{\circ}$ E, $16^{\circ}$ S & M1.8 & no & $95.7\pm 40.0$ & Bb \& Kd  \\
		&		&	  					&	&	& -$1684\pm 1914 $ &	\\
2003-10-29 & 20:37 & $9^{\circ}$ W, $19^{\circ}$ S & X10.0 & yes & $460$ & Bb \& Kd  \\
		&		&	  					&	&	&  $2423$ &	\\
2004-11-09 & 16:59 & $51^{\circ}$ W, $8^{\circ}$ N & M8.9  & yes & -- & Bb, Ka \& Kd  \\
&&&&&&\\
2005-05-06 & 16:03  & $28^{\circ}$ E, $9^{\circ}$ S & C8.5 & yes & $63.0\pm 15.8$ & Bb, Ka \& Kd \\
&&&&&&\\
2005-05-13 & 16:13  & $11^{\circ}$ E, $12^{\circ}$ N & M8.0 & yes & $36.3\pm 7.2$ & Ka \& Kd \\
		&		&	  					&	&	&   $153\pm 55.8$	 &	\\
2006-12-06 & 18:29 & $63^{\circ}$ E, $6^{\circ}$ S & X6.5  & yes & $851$ & Kb \& Kd \\
&&&&&&\\
2010-11-06 & 15:30 & $58^{\circ}$ E, $19^{\circ}$ S  & M5.4 & yes & $65.4\pm 4.8$ & Kc \& Kd \\
		&		&	  					&	&	&   $465\pm 206$ &	\\
2010-11-30 & 17:35  & $39^{\circ}$ E, $15^{\circ}$ N & None & yes & $51.0\pm 4.9$ 	 & Kd \\
\hline
  \end{tabular}
  }
 \end{center}
\vspace{1mm}

\end{table}

Between the initial parametrization of SCBs in 2005 and contemporary work completed in the past few years, several physical models have been put forth to phenomenologically describe where SCBs originate. Specifically, \cite[Balasubramaniam et al. (2006)]{2006ilws.conf...65B} found SCBs to be related to their host flare only in 65\% of cases studied and postulated that ``...SCBs are not a direct consequence of flares.'' This idea of flare independence differs with empirical models of \cite[Kirk et al. (2012a)]{2012ASPC..463..267K} and \cite[Pevtsov, Balasubramaniam, and Hock (2007)]{2007AdSpR..39.1781P} which show SCBs emerging from magnetic connectivity to the erupting active region. Both \cite[Balasubramaniam et al. (2006)]{2006ilws.conf...65B}  and \cite[Pevtsov, Balasubramaniam, and Hock (2007)]{2007AdSpR..39.1781P} assert that SCBs have stable stable mono-polar photospheric magnetic foot points while the work of \cite[Kirk et al. (2017)]{Kirk2017} finds a more complicated magnetic substructure. \cite[Kirk (2013)]{2013PhDT.......189K} and \cite[Kirk et al. (2014)]{2014ApJ...796...78K} found that SCBs are a type of localized chromospheric heating and ablation due to impacting coronal plasma. 

This work seeks to construct a cohesive narrative of SCBs' origin and evolution by collecting the many previous findings into one location. Since their discovery, only a handful of events have been positively identified to have SCBs. Table~\ref{tab1} lists all 14 events that have had SCBs analyzed over the last decade. This meager event list is mostly due to the way in which SCBs have been identified in all previous studies: with both high-resolution full-disk \Ha\ line-center images and complementary \Ha\ Doppler velocity maps. Synoptic, full-disk \Ha\ images are necessary to discover and analyze SCBs for two reasons. First, SCBs are fleeting -- lasting less than 7 minutes on average~(\cite[Kirk et al. 2014]{2014ApJ...796...78K}). Second, SCBs appear well outside of their parent eruption -- sometimes as far as 600 arcsec away~(\cite[Kirk 2013]{2013PhDT.......189K}). Other chromospheric absorption line profiles besides \Ha\ are all but guaranteed to show SCBs, but as of this writing, this has not been accomplished. Therefore all existing studies of SCBs have relied upon the Improved Solar Observing Optical Network~(\cite[ISOON; Neidig et al. 1998]{998ASPC..140..519N}) prototype telescope. ISOON was a semi-automated telescope producing 1.1 arcsec pixel, full-disk images of the Sun at a one-minute cadence and was decommissioned in 2011 and never replaced. With future observing of SCBs handicapped by the lack of appropreate instrumentation, the concatenation and review of SCB studies is paramount to creating a coherent understanding of these fascinating chromospheric events. 

\section{Observations: What Is an SCB and What Isn't It}

\begin{figure}
\begin{center}
 \includegraphics[width=3.5in]{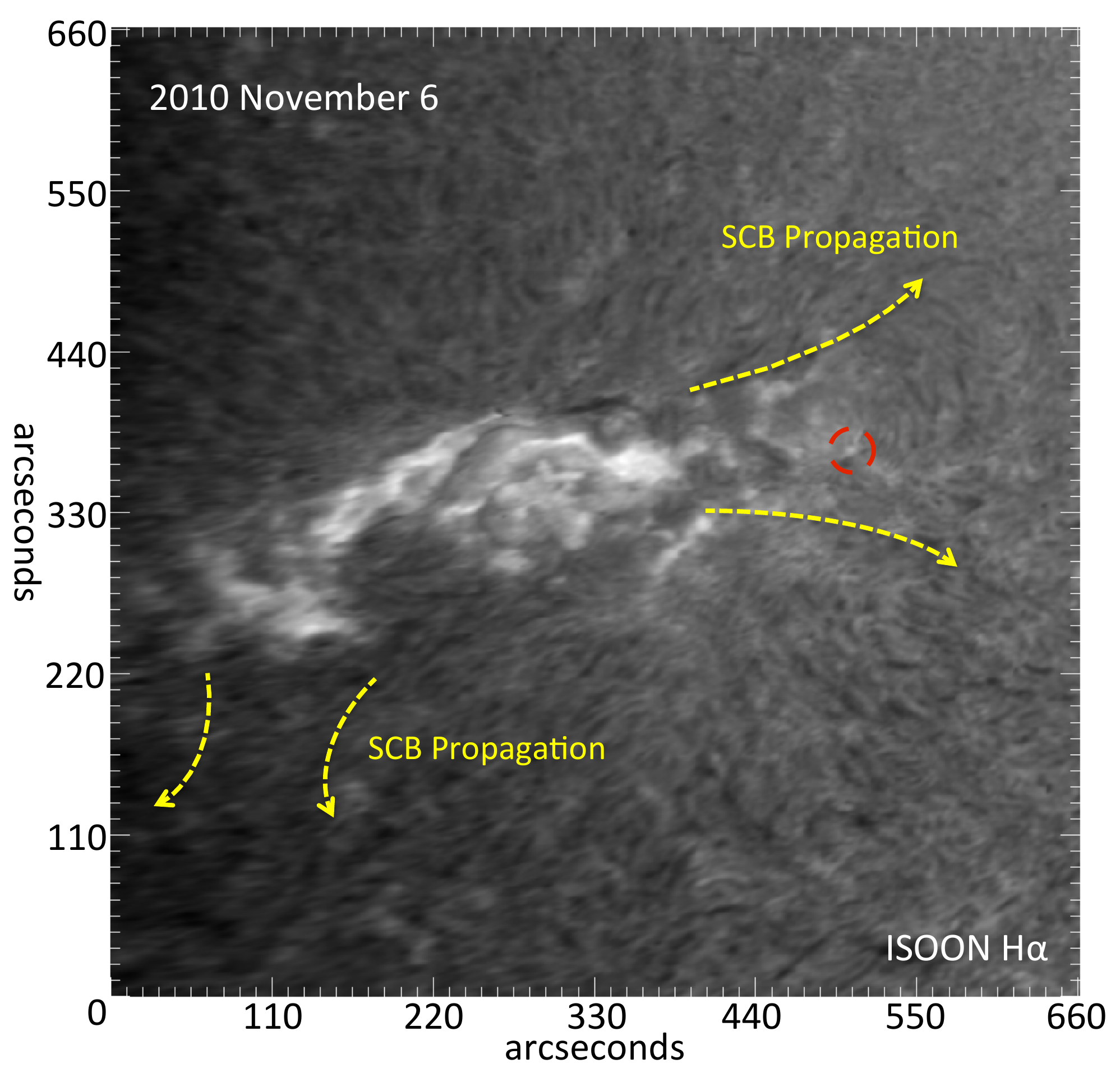} 
 \caption{An ISOON~\Ha\ image of a flare eruption that produced SCBs from 2010-11-06. The yellow dashed arrows show the approximate propagation direction and extent. Highlighted inside the red dashed circle is an example of a single SCB (further described in subsequent figures).}
   \label{SCBOverview}
\end{center}
\end{figure}

SCBs manifest as a series of spatially separated points in the chromosphere that brighten in sequence~(\cite[Balasubramaniam et al. 2005]{2005ApJ...630.1160B}). The sequential nature of the point brightening gives the appearance of a progressive traveling disturbance when a series of consecutive observations are animated. SCBs are observed in \Ha\ as sets of single and multiple trains of brightenings in association with a large-scale filament or flare eruption in the chromosphere or corona. In canonical two-ribbon flares, the loci of brightenings are observed as emerging predominantly along the axis of the flare ribbons rather than in the expansion directions of the ribbons. Figure~\ref{SCBOverview} shows an image of an SCB-producing flare with arrows indicating the observed propagation of SCBs. In more complex eruptions, SCBs regularly appear in all directions~(\cite[Kirk 2013]{2013PhDT.......189K}). Observationally, the appearance of SCBs are closely correlated with solar flares, coronal restructuring of magnetic fields, halo CMEs, EIT waves, and chromospheric sympathetic flaring~(\cite[Balasubramaniam et al. 2005, Kirk et al. 2017]{2005ApJ...630.1160B,Kirk2017}). 

SCBs are more closely related to the impulsive phase of the \Ha\ flare than any other part. SCB intensity curves are impulsive with a sharp peak and then a return to background brightness in the span of about 6.6 minutes~(\cite[Kirk et al. 2017]{Kirk2017}). Taken as an ensemble, SCBs routinely begin brightening about 30 minutes before the flare's peak-emission and return to background intensity about 20 minutes after and themselves have a peak intensity of 1.2 -- 2.5 times brighter than the average background intensity level~(\cite[Kirk 2013]{2013PhDT.......189K}). The associated flare, in contrast, can brighten more than an order of magnitude and remain above pre-flare levels for several hours. 

\begin{figure}
\begin{center}
 \includegraphics[width=2.8 in]{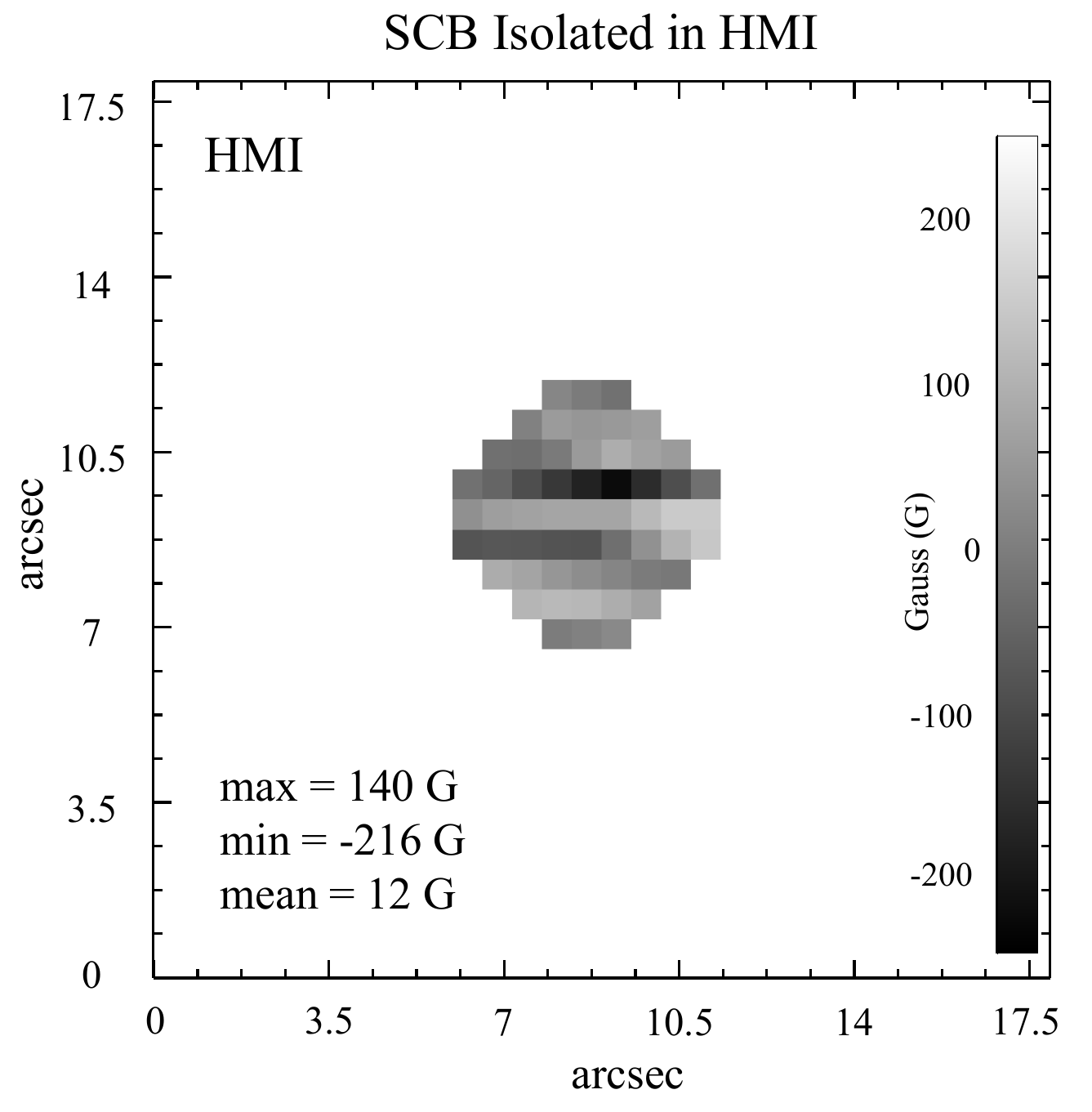} 
 \caption{The SCB highlighted in Figure~\ref{SCBOverview} isolated in HMI. The boundary of the SCB is determined by the flux in \Ha\ and then overlaid on the coincident photospheric magnetogram to isolate the magnetic signature of the SCB.}
   \label{SCBMag}
\end{center}
\end{figure}

Individual SCBs are physically small. They have an average radius of 2.4~Mm, which translates to only a couple of pixels in ISOON imagery. ISOON measurements may be under resolving their actual size. A study of SCBs that included complementary brightenting observed in the Solar Dynamics Observatory's (SDO) Atmospheric Imaging Assembly (AIA) imagery showed that SCBs appeared more compact in both space and time when observed with superior resolution~(\cite[Kirk et al. 2014]{2014ApJ...796...78K}). Figure~\ref{SCBEvol} shows an individual SCB isolated in \Ha, SDO AIA 304~\AA, 1600~\AA, and 1700~\AA, as well as the \Ha\ Dopplergram in which the substructure is visible in the SDO wavelengths is unresolved in \Ha.

\cite[Kirk et al. (2017)]{Kirk2017} measured the photospheric magnetic flux bounded by SCBs using the Helioseismic and Magnetic Imager (HMI). They found that the vector magnetic components of SCBs exhibit a clear nonzero result by having either distinct positive or negative direction with an average magnitude of $148 \pm 2.9$~G. Figure~\ref{SCBMag} shows an SCB that is isolated in an HMI magnetogram. The magnetic substructure observable in Figure~\ref{SCBMag} within the \Ha\ boundary of SCBs is statistically significant~(\cite[Kirk et al. 2017]{Kirk2017}). These findings suggest a magnetic footpoint of SCBs in the photosphere with a diameter significantly smaller than in that of the chromospheric brightening.

Several other types of compact transient brightening have frequently been observed in \Ha\ over the last century. SCBs are a distinct class of brightening and have distinguishable characteristics. Ellerman Bombs~(\cite[Ellerman 1917]{1917ApJ....46..298E}) are compact bright points of about the same size, however unlike SCBs they show little change in \Ha\ line-core intensity. Hyder Flares~(\cite[Hyder 1967]{1967SoPh....2...49H}) are progressive fronts of brightening in \Ha\ but only appear in the quiescent sun and never in the vicinity of flare eruptions. Moreton waves~(\cite[Moreton 1960]{1960AJ.....65U.494M}) are progressive fronts of brightening and appear with flare eruption. These waves also have a impulsive \Ha\ intensity line center like other SCBs. However, the Doppler profile of the Moreton wave is distinct. The Moreton wave begins with a positive Doppler shift, then there is an impulsive negative velocity shift followed by a shift back to a positive velocity before it decays to background levels (the essential quality of a progressive wave front), which is different than SCBs. The magnitude of Doppler velocities associated with the Moreton wave decays as a function of distance from the flare while there is no such trend in SCBs.

\section{SCB Dynamics, Propagation, and Heating}

When SCBs are viewed as an ensemble, they appear to progress outward from the flare. Each study previous study of SCBs have characterized this ensemble motion through various techniques (Table~\ref{tab1}). Progressive trains of SCBs are measured to propagate away from the flare with velocities in with a large dynamic range: from 35~\kms to speeds up to 260~\kms\ (speeds are also measured greater than 260~\kms\ with large uncertainties). However when the individual tracks of SCB kernels are examined, they do not show any progressive motion~(\cite[Kirk 2013]{2013PhDT.......189K}). The centroid of an SCB kernel randomly gyrates around within about six pixels of its starting location during the intensity enhancement. Although the ensemble of SCBs demonstrate a sequence of point brightening, giving the appearance of a progressive traveling disturbance, the bright emission of an individually measured SCB does not follow the disturbance and remains in the same location. Similar to a wave, the medium in which SCBs are measured remains laterally undisplaced with the apparent propagation of the brightenings.

\begin{figure}
\begin{center}
 \includegraphics[width=3.2in]{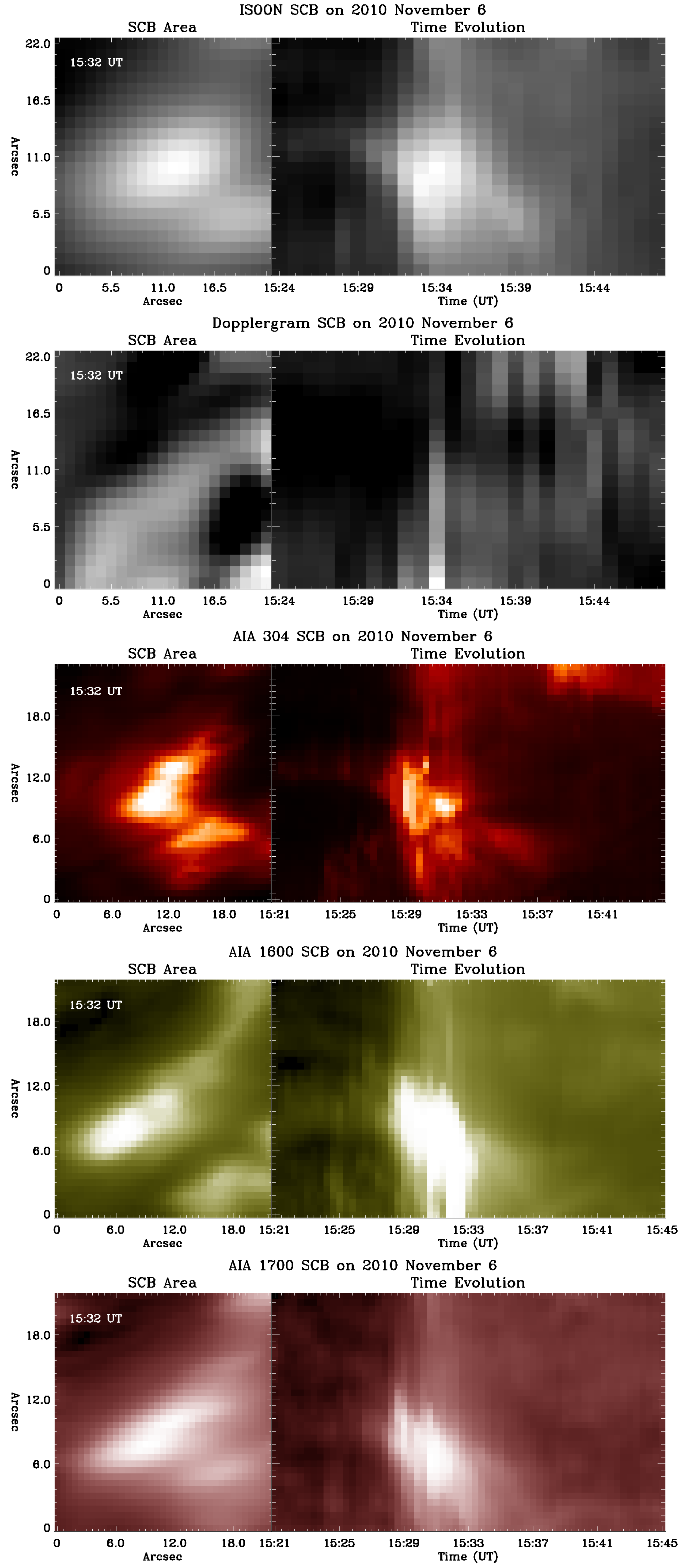} 
 \caption{Images and temporal evolution of the SCB highlighted in Figure~\ref{SCBOverview} in \Ha\ and SDO AIA 304, 1600, and 1700 channels. For each data set, the left side image shows the isolated SCB from November 6 (a light curve of this event is also shown in Figure~\ref{SCBDop} as an example of a type II SCB). The right side extracts a column in the core of the isolated SCB (at 11 arcsecs) and shows its time evolution.}
   \label{SCBEvol}
\end{center}
\end{figure}

\begin{figure}
\begin{center}
 \includegraphics[width=4.7 in]{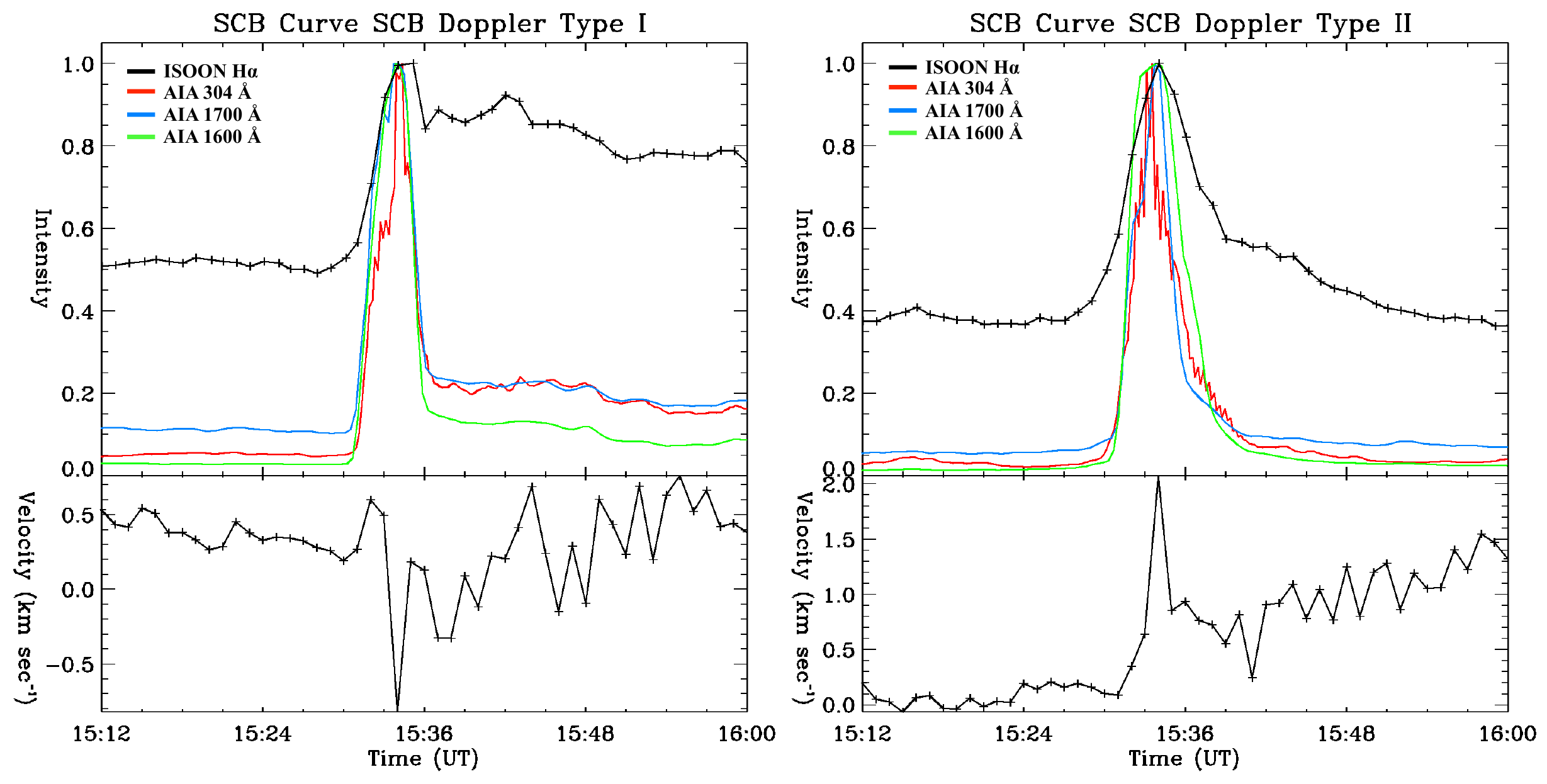} 
 \caption{Examples of SCBs of type I and type II. The top plot shows the normalized intensity curves:~\Ha\ in black, AIA 1600 channel in green, 1700 channel in blue, and 304 channel in red. The bottom panel plots the measured \Ha\ Doppler velocity. The type II SCB is the same SCB as highlighted in Figure~\ref{SCBOverview}.}
   \label{SCBDop}
\end{center}
\end{figure}

\cite[Kirk et al. (2012b)]{2012ApJ...750..145K} define three different types of SCBs based upon their corresponding Doppler signature. A type I SCB has an impulsive intensity profile and an impulsive negative Doppler profile that occurs simultaneously or a few minutes after the peak brightening (Figure~\ref{SCBDop} left panel). In this context, a negative velocity is associated with motion away from the observer and into the Sun. A type II SCB has a positive Doppler shift perturbation that often lasts longer than the emission in the \Ha\ intensity profile (Figure~\ref{SCBDop} right panel). The timing of both are nearly coincidental. A type III SCB demonstrates variable dynamics (not shown). It has a broad \Ha\ intensity line center with significant substructure.  A type III SCB begins with a negative Doppler profile much like a type I. Before the negative velocity perturbation can decay back to continuum levels, there is a dramatic positive velocity shift within three minutes with an associated line center brightening.  In all types of SCBs the typical magnitude of the perturbation perpendicular to the solar surface of the Doppler velocity is between 1 and 5 \kms\ from the pre-SCB \Ha\ velocity.

SCBs have an impulsive intensity evolution through time -- they brighten to peak intensity and return to background levels over the span of a few minutes.   Figure~\ref{SCBEvol} shows an example of a type II SCB in ISOON~\Ha, SDO AIA 304~\AA, 1600~\AA, and 1700~\AA, as well as the \Ha\ Dopplergram. Figure~\ref{SCBEvol} also readily demonstrates the differing resolutions in both space and time between the data sets. A typical SCB observed in AIA also has a peak intensity delayed by about a minute as compared to ISOON~(\cite[Kirk et al. 2012b]{2012ApJ...750..145K}). This delay is more pronounced in the 1600~\AA\ and 304~\AA\ images, which are primarily made up of emission that forms in the high chromosphere, than the 1700~\AA\ image, which forms from plasma in the mid to low chromosphere. These measurements imply that SCBs are formed in the mid-chromosphere and propagate vertically upward toward the transition region and downward toward the photosphere.

The simplest conceptual idea of an SCB is a volume of heated chromospheric plasma. In a simple radiative cooling model, \cite[Kirk et al. (2014)]{2014ApJ...796...78K} calculate an SCB cooling time of $t_{\rm cool}\simeq 10^{-1}$~s. This simplistic model is far from physical because the chromosphere is incompletely ionized; electrons are not singularly responsible for the temperature of an SCB; thermal conductivity is not infinite; and other heat transfer processes are ignored. This model does provide a lower bound to the cooling time of SCBs. \cite[Carlsson \& Stein (2002)]{2002ApJ...572..626C} use a non-LTE treatment of hydrogen, calcium, and helium, and they account for both radiative and collisional processes at a range of densities and column mass. They find a chromospheric relaxation time for hydrogen at a height of 2~Mm above the photosphere to be $t_{\rm relax} \simeq 10^{2.5}$~s.  Separately, \cite[Giovanelli (1978)]{1978SoPh...59..293G} calculates a chromospheric relaxation time: $t_{\rm relax}\simeq 10^2$~s in the low chromosphere up to $t_{\rm relax}\simeq 10^{2.6}$~s in the upper chromosphere. In the simplistic radiative cooling model as well as two more careful calculations, chromospheric hot spots should dissipate (through kinetic and radiative processes) in a couple of minutes. The relaxation time in the upper chromosphere is comparable to both the 4.5 minute median SCB duration in SDO AIA 304~\AA\ and the 6.6 minute duration in \Ha. The duration of SCBs compared to the local relaxation time implies that some longer duration SCBs are actively heated over a significant portion of their lifetime, while the shorter-lived SCBs can be caused by one isolated heating event. 

Returning to a prototypical SCB of average radius, a reasonable estimate for the duration of SCB heating is about $t_{\rm heating}=10^{2.6}$~s, which is the median duration of the \Ha\ intensity enhancement. If the chromosphere is heated at a rate of $\Lambda=4.5 \times 10^9$~erg~g$^{-1}$~s$^{-1}$~(\cite[Anderson \& Athay 1989]{1989ApJ...346.1010A}), then the total energy required to heat a single SCB is 
\begin{equation}
E_{\rm SCB} \simeq \Lambda V n_e m_p t_{\rm heating} \simeq 5 \times 10^{25}\ {\rm erg}, 
\end{equation}
where $V$ is the volume of the SCB, $m_p $ is the mass of a proton, and assuming a neutral plasma. Given a typical mid-size flare, SCBs account for as much as $\approx 0.01\%$ of the flare energy budget~(\cite[Kirk et al. 2014]{2014ApJ...796...78K}). SCBs are an insignificant portion of the total energy released in a solar flare and given a lack of correlation between number of SCBs and flare class, they likely are not directly heated by the flare reconnection~(\cite[Kirk et al. 2017]{Kirk2017}). 

\section{Physical Origin}
Each new analysis of SCBs refines the narrative of their physical origin. \cite[Balasubramaniam et al. (2006)]{2006ilws.conf...65B} find SCBs to be related to their host flare only in 65\% of cases studied and postulate that ``...SCBs are not a direct consequence of flares.'' In this way, SCBs are unlike flares ribbons in that they are a secondary effect of flare reconnection events. SCBs are more closely related to the release of an associated CME. This view of SCBs as causally connected to CME release is bolstered by the finding that flares releasing vastly different amounts of energy (from no X-ray signature to an X10.0 flare) have SCBs with similar physical traits (Table~\ref{tab1}). Since SCBs have been observed to first appear before the onset of \Ha\ emission in a flare, during the impulsive phase, or after the peak intensity, an associated CME is most likely to be the process driving the appearance of SCBs rather than flare reconnection~(\cite[Kirk et al. 2017]{Kirk2017}). Observations of SCBs related to flare and filament eruptions can give an indication of an emerging CME a couple hours before it is detected a coronagraph. 

From the relationships between SCBs and CMEs and regional magnetic field extrapolations, \cite[Kirk et al. (2017)]{Kirk2017} suggest that SCBs originate in quiescent coronal magnetic loops above a chromospheric flare. These loops are forcibly disturbed when a flare or filament erupts. As these coronal tethers reconfigure into a new equilibrium, trapped plasma in the tethers is now free to cascade into the chromosphere causing SCBs. This model is affirmed by the finding that 86\% (12 out of 14 in Table~\ref{tab1}) of SCB events in this study accompanying subsequent visible CMEs, also validating the conclusions of~\cite[Balasubramaniam et al. (2006)]{2006ilws.conf...65B}. Therefore the progressive trains of SCBs observed are coronal in origin yet appear in the chromosphere. SCBs are indicative of the pre-flare surrounding magnetic topology rather than the flare itself; reminiscent of the tethers identified in breakout reconnection~(\cite[Antiochos et al. 1999]{999ApJ...510..485A}). 

Individual SCBs are not energetic enough to be isolated examples of chromospheric evaporation. Like chromospheric evaporation, an incident beam of high energy particles heats the SCB volume of plasma to $T \simeq 10^5$~K. The heated SCB adiabatically expands vertically upwards and downwards, confined by the magnetic flux tube, at speeds much less than the local sound speed but does not fully ablate from the chromosphere~(\cite[Kirk et al. 2014]{2014ApJ...796...78K}). The chromospheric heating leading to an SCB continues to persist over a significant portion of the SCB's lifetime. By estimating the energy required to heat SCBs to be $E_{\rm SCB} \simeq 5 \times 10^{25}\ {\rm erg}$, it is unlikely that SCBs have enough energy to evaporate the heated chromospheric material back into the corona. The heated material in SCBs pushes into the upper-chromopshere/transition region and then collapses back down into the mid-chromosphere after cooling.

A couple of studies in the past couple of years have added additional constraints on the physical nature of SCBs. \cite[Gilbert et al. (2013)]{2013ApJ...776L..12G} tracked coronal material reentering the chromosphere from a failed prominence eruption. They find the impact area of these events to have a radius of $16-228$~Mm, which is one to two orders of magnitude larger than studied SCBs. These impact sites also total radiated energy of $10^{25} - 10^{26}$~erg and total kinetic energies estimated to be an order of magnitude higher. This results in a radiated energy density of $10^{4} - 10^{7}$~erg cm$^{-2}$, which is at least an order of magnitude smaller than an SCB's radiated energy density ($\simeq 10^{8}$~erg cm$^{-2}$ for an average SCB). This means the material that heats SCB's is accelerated to velocities beyond what \cite[Gilbert et al. (2013)]{2013ApJ...776L..12G} measure. 

A high-resolution study of an erupting M6.5 flare by \cite[Jing et al. (2016)]{2016NatSR...624319J} notes brightening at several impact cites of coronal rain at a much finer scale than any SCBs previously studied. Unlike SCBs, these brightening appear at the footpoints of visually discernible post-flare loops and occur simultaneously with the impact of dense plasma - a compact analogue to the features studied by \cite[Gilbert et al. (2013)]{2013ApJ...776L..12G}. These coronal rain compact brightenings move in a sequential pattern and trail behind recently formed flare ribbon's path of progression~(\cite[Jing et al. 2016]{2016NatSR...624319J}). Visually, they behave as if they are an `echo' of the flare eruption but with an energetically negligible magnitude. It is difficult to conclusively say if these brightenings have the same origin as SCBs, but from the formation location alone, they seem to be distinct features. 

\section{Future Directions of Research into SCBs}

SCBs remain a tantalizing chromospheric feature that are understudied and may have significant relationships to other eruptive phenomena. In the last five years, breakout reconnection~(\cite[Antiochos et al. 1999]{1999ApJ...510..485A}) has been used to explain both SCBs~(\cite[Kirk et al. 2017]{Kirk2017}) and the EUV late phase (ELP) of solar flares~(\cite[Hock 2012, Woods et al. 2011]{Hock2012,2011ApJ...739...59W}). Both SCBs and ELP emission were observed during the 2010 November 6 solar flare (see Table~\ref{tab1} and Figure~\ref{SCBOverview} for details on this event).  In fact, the SCBs appear to be collocated with the secondary post-flare loop system that produce the ELP. Future research will detail at the relationship between eruptions with SCBs and ELP flares to determine whether SCBs could be used a predictors of ELPs.

\cite[Kuridze et al. (2013)]{2013A&A...552A..55K} also studied the eruption on November 6 in to characterize coronal material draining out of a partial filament eruption. They track filamentary material falling into the sun and the development of a strong helical twist into the erupted filament.  The evolution of the erupting filament supports the magnetic breakout model of eruption and is attributed to the observed evolution of the event~(\cite[Kuridze et al. 2013]{2013A&A...552A..55K}). They also surmise that a large-scale, closed, overlying magnetic loop arcade confined the eruption. Both of these findings, breakout reconnection and large overlying loop structure, support the model of SCB formation. Further work to correlate filamentary twist and draining material to SCBs observed in other filament eruptions is necessary to fully explore the relationship between filaments and SCBs. 

In recent work by \cite[Zhang et al. (2017)]{2017A&A...598A...3Z}, they document large-scale coronal dimming that occurs in advance of a flare eruption but is coincident with a CME release. The pre-flare dimmings are found to originate from density depletion that is a result of the gradual expansion of a regional coronal loop system during the slow rise of the erupting flux rope. This type of event is also likely to produce SCBs given the large-scale reorganization of a coronal loops and provides a means to conclusively determine if SCBs occurring in the impulsive phase of their host flare are likely originate from a CME eruption. Mapping the spatial and temporal relationship between SCBs and coronal dimming is likely to produce a more concrete model of when exactly the progenitor of SCBs is triggered. 

Lasting only a few minutes, SCBs are fleeting indicators of the solar flare environment at the time of flare eruption. It would be easy to dismiss these compact bright points as just another minor effect of a flare or CME eruption. This attitude would be short sighted given the potential connection of SCBs to larger-scale phenomena such as ELP emission, filament draining, and coronal dimming. Future studies of SCBs should focus on placing them into a greater context with coronal eruptions to establish the large-scale dynamic relationship between the chromosphere and corona. Currently, SCBs are uniquely identified by their Doppler intensity profile in \Ha\ images. ISOON has been the only solar imaging telescope that regularly recorded full-disk off-band images in the \Ha\ line with suitable resolution to capture SCBs. Unfortunately this telescope was decommissioned in 2011 and as a direct result the ongoing study of SCBs has also stagnated. 

\acknowledgments
The authors would like to acknowledge the Thomas Metcalf Lecturer Program from the American Astronomical Society's Solar Physics Division for their generous travel support to attend and present at the IAU Symposium 327.

\end{document}